\documentclass[pre,aps,amsmath,amssymb,showpacs,floatfix,12pt]{revtex4-1}
\usepackage{graphicx}
\usepackage{graphicx}
\usepackage{graphics}
\usepackage{psfrag}

\newcommand{\bq}{\begin{equation}}
\newcommand{\eq}{\end{equation}}
\newcommand{\ba}{\begin{array}}
\newcommand{\ea}{\end{array}}

\renewcommand{\v}{\mathrm{v}}

\begin{document}

\title{A refined empirical stability criterion for nonlinear Schr\"odinger solitons under spatiotemporal
forcing}

\author{Franz G.\ Mertens}
\email{Franz.Mertens@uni-bayreuth.de} \affiliation{Physikalisches
Institut, Universit\"at Bayreuth, D-95440 Bayreuth, Germany}

\author{Niurka R.\ Quintero}
 \email{niurka@us.es}
\affiliation{Departamento de F\'\i sica Aplicada I, E.U.P.,
Universidad
de Sevilla, c/ Virgen de \'Africa 7, 41011 Sevilla, Spain}

\author{I. V. Barashenkov}
\affiliation
{Department of Mathematics, University of Cape Town, Rondebosch 7701,
South Africa}

\author{A. R. Bishop}
\affiliation
{Los Alamos National Laboratory,  Los Alamos, NM 87545, USA}

\date{\today}


\begin{abstract}
We investigate the dynamics of travelling oscillating solitons of the cubic 
NLS equation under an external spatiotemporal forcing of the form 
$f(x,t) = a \exp[iK(t)x]$. For the case of time-independent forcing a stability 
criterion for these solitons, which is based on a collective coordinate theory, 
was recently conjectured. We show that the proposed criterion has a limited 
applicability and present a refined criterion which is generally applicable, 
as confirmed by direct simulations. This includes more general situations 
where $K(t)$ is harmonic or biharmonic, with or without a damping 
term in the NLS equation. The refined criterion states that the soliton will 
be unstable if the ``stability curve''  $p(\v)$, where $p(t)$ and $\v(t)$ are 
the normalized momentum and the velocity of the soliton, has a section with a 
negative slope. Moreover, for the case of constant $K$ and zero damping 
we use the collective coordinate solutions to compute a ``phase portrait''  
of the soliton where its dynamics is represented by two-dimensional 
projections of its trajectories in the four-dimensional space of 
collective coordinates. 
We conjecture, and confirm 
by simulations, that the soliton is unstable if a section of the 
resulting closed curve on the portrait has a negative sense of rotation.
\end{abstract}

\maketitle

\section{Introduction}
The externally driven, nonlinear Schr\"odinger (NLS) equation arises 
in many applications, for example charge density waves \cite{kaup}, 
long Josephson junctions \cite{LJJ}, optical fibers \cite{C,D,cohen} 
or plasmas driven by rf fields \cite{E}. We use the NLS in the form 
\begin{equation}
\label{aa}
i u_t + u_{xx} + 2|u|^2 u + \delta u = R[u(x,t);x,t],   
\end{equation}
with the perturbation 
\begin{equation}\label{aa1}
R = f(x,t)- i \beta u(x,t),
\end{equation}
where $f(x,t)$ is a direct (external) driving force  
and the term with $\beta$ accounts for dissipation.  
Different forms of the driving force were considered: e.g., 
ac driving  $f = \epsilon \exp(i \omega t)$ \cite{kaup,Bar1,Bar3} or driving 
by a plane wave $f = \epsilon \exp[i(kx-\omega t)]$ \cite{cohen,vyas}. 
Moreover, $f = \epsilon \exp[ig(x,t)-i \omega t]$, where $g$ is a 
function of $x-vt$, was considered, but no localized solutions were 
discussed \cite{vyas}. 

The present paper continues the analysis 
\cite{fgm} of the soliton dynamics under the spatiotemporal 
driving
\begin{equation}\label{aa2}
f(x,t) = a e^{i K(t) x}.  
\end{equation}
A discrete version of Eq.\ (\ref{aa}) was used to model nonlinear optical 
waveguide arrays, 
in which discrete cavity solitons
can be excited \cite{oll5}. In that application, $\delta$ is the cavity detuning parameter, 
and $f(x,t)$ is replaced with   
$f_n (t) = a \exp(i \phi_{in} n)$, where $n$ numbers the resonators and $\phi_{in}(t)$ is the incident angle of 
the laser pump light. A biharmonic function $\phi_{in}(t)$ was used in order to 
generate a ratchet effect \cite{gorbach}. In the present paper we also 
obtain a ratchet effect by using a biharmonic driving (Section \ref{sec5}). 
This is interesting because there are only a few reports on ratchets with 
nontopological solitons \cite{gorbach,poletti,rietmann}; 
most of the literature concerns ratchets with topological 
solitons, e.g \cite{sazo,prl2003,usti,franz,franz1}.

In Ref.\ \cite{fgm}, Eq.\ (\ref{aa}) was simulated using the 1-soliton solution of the 
unperturbed NLS as the initial condition.
The soliton's position, velocity, amplitude and phase 
 served 
as parameters of the initial
condition (IC).      

In the case of zero damping and time-independent, spatially periodic 
driving of the form $f(x) = \exp(i K x)$ the resulting solitons were 
observed \cite{fgm} to display periodic oscillations of their  
positions, velocities, amplitudes and phases. 
Although the driving force has 
zero spatial average, the soliton's net motion is 
\emph{unidirectional}. (This contrasts with the case of a 
perturbation $V(x) u$ with  
periodic $V(x)$,  
where the soliton performs oscillations about a minimum of $V(x)$ \cite{scharf}). 

A large number of sample points in the parameter space $(a, K, \delta)$ 
were examined by varying the initial amplitude $\eta_{0}$, with the other 
initial conditions kept fixed. 
The initial configuration was seen to evolve into a
stable soliton only when $\eta_0$ was taken from one of the ``stability windows''. 
For $\eta_0$ outside the 
stability windows, the solitonic  
initial condition was observed to decay or break into two or more fragments which would 
subsequently decay \cite{fgm}.

As a first step towards understanding the observed dynamics of solitons, 
the authors of Ref.\ \cite{fgm} 
proposed an empirical stability criterion 
based on a collective coordinate (CC) description. The collective coordinates analysis
produces a set of coupled nonlinear ODEs for the soliton's position $q$, 
amplitude $\eta$, normalized momentum $p$ and phase $\Phi$.  
An approximate solution of this dynamical system is given by trigonometric functions 
and can be obtained explicitly, except when the initial condition $\eta_0$ is 
chosen near one of the stability boundaries. In the latter case the collective coordinates 
equations had to be analysed numerically and their solutions 
were found to be highly anharmonic. 

%

We have positively tested the predictions of the proposed stability criterion by simulations 
(numerical solutions of the full NLS Eq.\ (\ref{aa})) for many classes of initial conditions. 
However, the tests are negative, if the initial momentum is too large, i.e. 
$p_0 > K$ for positive $K$ (or $p_0 < K$ for negative $K$). 
In this paper we therefore conjecture a refined stability criterion, 
which we show makes correct predictions not only for the case $K = constant$ 
(with all classes of initial conditions), but also for harmonic and biharmonic 
$K(t)$. The new criterion is a sufficient condition which states that the soliton will be unstable 
in simulations, if the ``stability curve''  $p(\v)$ has a branch with 
negative slope. This curve is obtained 
as a parametric plot of the \emph{normalized} momentum 
\begin{equation}\label{ab}
p=\frac{P(t)}{N(t)},
\end{equation}
of the soliton versus its velocity,  
\begin{equation}\label{ac}
\v=\dot{q}(t). 
\end{equation}
Here $P= 4 \eta p$ is the canonical momentum of the soliton, $N=4 \eta$ 
is the norm which is canonically conjugated to the 
soliton's phase $\Phi(t)$, see Section \ref{sec2}. In the old criterion 
\cite{fgm} the stability curve was defined as $P(\v)$. At the end of Section 
\ref{sec3} we present an example which demonstrates 
by an analytical calculation that the normalized momentum $p$, 
instead of the canonical momentum $P$, has to be used for the stability criterion.

It is important to emphasize that the soliton's stability or instability 
is judged not on 
the basis of the stability of solutions to the collective coordinates equations. 
(The latter are stable in most cases).  
The soliton's stability is rather decided on the basis 
of  some of its properties which are captured by the $p(\v)$ curve 
of the corresponding collective coordinates solutions. 
The proposed empirical criterion reproduces the numerically observed 
positions of the stability windows to an accuracy 
of better than $1\%$, despite the complexity of the stability diagram in the parameter space  
\cite{fgm}.

The stability criteria for the homogeneous 
(translation invariant) NLS equation available in the literature 
are restricted to 
(a) bright solitons, 
i.e. solutions decaying to zero at the spatial infinities, with time dependencies  
of the form $e^{i \Lambda t}$ (and those reducible to this form by a Galileian 
transformation); 
(b) traveling dark solitons, i.e. solutions approaching  nonzero constant values as $x \to \pm \infty$. 
The criteria 
are insensitive to the particular form of the nonlinearity as long as it is conservative and $U(1)$-invariant, i.e.  
as long as the equation does not 
include any damping or driving terms.   


In the case of the bright solitons of the form 
$u(x,t)= u_s(x) e^{i \Lambda t}$, the Vakhitov-Kolokolov
criterion states that if the corresponding energy Hessian has only 
one negative eigenvalue, then 
the soliton is stable if  $dN/d \Lambda>0$ and unstable otherwise 
\cite{vk,w,A}. Here $N= \int |u|^2dx$; depending on the physical context, 
$N$ is referred to as 
the number of particles contained in the soliton or the total power of the 
optical beam. 
(See also \cite{B} for the energy-versus-number of particles 
formulation of this criterion.)

In the case of dark solitons of the form $u(x,t)=u(x-{\tilde V} t)$, with $|u|^2 \to \rho_0$ as $|x| \to \infty$,
a similar criterion 
\cite{BKK,BP,igor2,PKA} involves the (renormalised) field momentum, 
\begin{equation}\label{ad}
\tilde{P} = \frac{i}{2} \int  (u_{x}^* u - u_x u^*)
\left(1-\frac{\rho_0} {|u|^2}\right) dx.
\end{equation}
The dark soliton travelling at the constant velocity ${\tilde V}$ is stable 
if $d{\tilde P}/d{\tilde V}<0$
and unstable otherwise.

Some parts of the stability analysis of the travelling dark solitons \cite{igor2} can be carried over to
the case of the travelling solitons of the NLS with a driving term. Namely, 
one can show \cite{igor1} that a linearised eigenvalue crosses from the negative to the positive
real axis at the value ${\tilde V}$ where $d{\tilde P}/d{\tilde V}=0$.
The sign of the derivative $d{\tilde P}/d{\tilde V}$ required for stability depends on the 
type of the soliton; some classes of solitons require 
$d{\tilde P}/d{\tilde V}<0$, 
whereas other classes are stable when $d{\tilde P}/d{\tilde V}>0$. (An additional complication 
is the presence of oscillatory instabilities where two eigenvalues
collide on the imaginary axis and acquire opposite real parts. The oscillatory
instabilities do not affect the sign of $d{\tilde P}/d{\tilde V}$.)

In these analyses, each point of the curve $\tilde{P}(\tilde{V})$
represents a soliton traveling at a particular constant velocity $\tilde{V}$;
therefore the curve is a characteristic of the whole family of solitons.
The values of ${\tilde V}$  where $d\tilde{P}/d\tilde{V}=0$ break the family
into parts with different stability properties.
 In contrast to this, each oscillatory solution of the collective coordinates equations  
\cite{fgm} has its own, individual, $p(\v)$-curve
 the whole of  which is traced
 periodically in time. The shape of this  curve  determines whether the corresponding soliton is stable
 or not.  

The present paper has several goals: 
First, we propose a refined stability criterion. 
Second, we study the internal structure of the instability 
regions. We will demonstrate 
that these regions consist of subregions characterized by 
instabilities of different types. The 
existence of the subregions 
will be predicted by the analysis of the reduced dynamical system and 
confirmed by direct 
simulations of the full PDE (Section \ref{sec3}). In obtaining the reduced 
dynamical system we modify the original collective coordinates approach of Ref. \cite{fgm} (Section \ref{sec2}). 
In addition to producing bounded trajectories 
(a property essential for the stability analysis), the modified approach provides  
a much easier derivation of the canonical soliton momentum and 
the Hamilton function in terms of the canonical variables (Section \ref{sec2}). 

Third, we demonstrate that a certain ``phase portrait'' of the 
soliton on the complex plane can be used as an alternative stability diagnostic (Section \ref{sec3}). 
However, the phase portrait requires the phase of the soliton to be periodic in time. 
This can be achieved by the above mentioned modification of the original collective coordinates approach 
\cite{fgm} in which the phase was \emph{not} periodic, in contrast to the other three collective coordinates. 

Finally, we explore the applicability of 
our refined stability criterion to inhomogeneous forcings of the form 
$f(x,t) = a \exp(i K(t) x)$ in Eq.\ (\ref{aa2}). 
We will start with a harmonically varying $K(t)$, with and without 
the damping term in the right-hand side of (\ref{aa}) (Section \ref{sec4}). 
After that, in Section \ref{sec5},  
we will consider a \emph{biharmonic} $K(t)$ with a 
broken temporal symmetry. (The temporal symmetry breaking will accompany the breaking 
of the
spatial symmetry by the inhomogeneous driving.)  

\section{Modified Collective Coordinate Theory}\label{sec2}
The one-soliton solution of the unperturbed NLS is given by \cite{kivmal}   
\begin{equation}\label{ba}
u(x,t) = 2 i \eta \, \mathrm{sech} [2 \eta (x- \zeta)] e^{-i(2 \xi x + \phi)},
\end{equation}
where $\eta$ and $\xi$ are real parameters ($\eta >0$); $\zeta = \zeta_{0}-4 \xi t$ gives 
the coordinate of the soliton's center, 
and $\phi=\phi_{0}+(4 \xi^2-4 \eta^2-\delta) t$ is the soliton's phase. The
collective coordinates theory of Ref.\ \cite{fgm} assumed that for sufficiently small perturbations $R$ in 
Eq.\ (\ref{aa1})   
the soliton shape and dynamics can be described, approximately, 
by Eq.\ (\ref{ba}), where $\eta(t)$, $\xi(t)$, $\zeta(t)$ and $\phi(t)$ are functions 
of time.

We now show that the following modification of this ansatz \cite{modi,quintero:2010} 
provides a considerable  
improvement of the collective coordinates theory of Ref. \cite{fgm}: 
\begin{equation}\label{bd}
u(x,t) = 2 i \eta \, \mathrm{sech} [2 \eta (x- q)] e^{i[p(x-q) - \Phi]},
\end{equation}
by setting $- 2 \xi=p$, $\zeta=q$, and $\phi=\Phi-2 \xi \zeta=\Phi+pq$. Here only the last 
replacement is essential for the above mentioned improvement of the 
collective coordinates theory. 
The four collective coordinates equations of Ref.\ \cite{fgm} are replaced with 
\begin{eqnarray}
\label{be1}
\dot{\eta} &=& - 2 \beta \eta - \frac{\pi}{2} a \, \mathrm{sech} A \cos B,\\
\label{be2}
\dot{q} &=& 2 p + \frac{\pi ^2}{8} \frac{a}{\eta ^2} 
\mathrm{sech}
A \tanh A \sin B, \\
\label{be3}
\dot{p} &=& -2 a A \, \mathrm{sech}
 A \cos B, \,\\
\label{be4}
\dot{\Phi} + p \dot{q} &=&  p^2 - 4 \eta ^2 - \delta + 
\frac{\pi}{2} \frac{a A}{\eta} \mathrm{sech}
A \tanh A \sin B,
\end{eqnarray}
with
\begin{eqnarray}
\label{bf1}
A(t) &=& \frac{\pi}{4 \eta(t)} [K(t) -p(t)], \\
\label{bf2}
B(t) &=& \Phi(t) + K(t)q (t).
\end{eqnarray}
The new formulation has the following advantages

1. Consider the Lagrangian for Eqs.\ (\ref{be1})-(\ref{be4}):   
\begin{equation}\label{bg}
L= 4 \eta \dot{\Phi}+ 4 \eta p \dot{q}-4 \eta p^2+\frac{16}{3} \eta^3+ 4 \delta \eta - 
2 \pi a \, \mathrm{sech} A \sin B. 
\end{equation}
The momentum conjugate to the phase $\Phi$ is 
\begin{equation}\label{bh}
\frac{\partial L}{\partial \dot{\Phi}} = 4 \eta, 
\end{equation}
which is equal to the norm $(\int |u|^2 \, dx)$ of the waveform 
(\ref{bd}). 
The momentum conjugate to the soliton's position is   
\begin{equation}\label{bi}
\frac{\partial L}{\partial \dot{q}} = 4 \eta p.  
\end{equation}
The advantage of the new formulation is that this is equal to the field momentum of the configuration 
(\ref{bd})  
\begin{equation}\label{bi1}
P = \frac{i}{2} \int_{-\infty}^{+\infty} (u_{x}^{*} u - u_{x} u^{*}) dx,   
\end{equation}
whereas in Ref. \cite{fgm} 
the second canonical momentum was defined by 
$\partial L/\partial \dot{p} = -4 \eta q$
which did not have any obvious physical 
interpretation. 

If the dissipative term $-i \beta u$ in (\ref{aa1}) has a nonzero coefficient, we have 
to use the generalised Euler-Lagrange formalism with the dissipation function 
\begin{equation}\label{bk1}
F= i \beta \int_{-\infty}^{+\infty} (u u_{t}^{*}- u^{*} u_{t}) dx.
\end{equation}
Substituting (\ref{bd}) in (\ref{bk1}), we obtain 
\begin{equation}\label{bk}
F=-8 \beta \eta (\dot{\Phi}+p \dot{q}).
\end{equation}
The generalised Euler-Lagrange equations are  
\begin{equation}\label{bj}
\frac{d\,}{dt} \frac{\partial L}{\partial \dot{\Psi}} - \frac{\partial L}{\partial \Psi}=\frac{\partial F}{\partial \dot{\Psi}},
\end{equation}
where $\Psi$ represents each of the four collective coordinates $\eta$, $q$, $p$, and $\Phi$. 

2. Since $P= 4 \eta p$ is the canonically conjugate momentum for $q$, 
the Legendre transform to the canonical Hamiltonian is easily performed: 
 $H=N \dot{\Phi}+P \dot{q}-L$. This gives  
\begin{equation}\label{bl}
H = \frac{1}{N} P^{2} - \frac{1}{12} N^{3} - \delta N+ 2 \pi a \, 
\mathrm{sech} A \sin B.
\end{equation}
In Ref. \cite{fgm} this Hamiltonian could only be obtained via a canonical transformation. 


When the forcing $f(x) = a e^{i K x}$ is time-independent and 
 damping $\beta=0$, the collective coordinates $\eta$ and $p$ 
perform periodic oscillations, whereas $q(t)$ and $\Phi(t)$ are 
given by periodic functions superimposed over linearly growing functions of $t$. 
In contrast to this, the variable $\phi=\Phi+p q$  
 used in Ref.\ \cite{fgm} will obviously exhibit oscillations 
with a (linearly) growing amplitude. 

\section{Time-independent, spatially periodic force} \label{sec3}

Our approach consists in the numerical solution of the collective coordinates equations  
(\ref{be1})-(\ref{be4}) for representative values of the parameters and the 
initial conditions $\eta_{0}$, $q_{0}$, $p_{0}$ and 
$\Phi_{0}$. Each collective coordinates orbit is then used to compute $p(t)$ and 
$\v(t)$ in Eqs.\ 
(\ref{ab})-(\ref{ac}) and to plot $p$ against $\v$. If some part of this 
``stability curve'' has a 
negative slope, we 
predict that the soliton will become unstable in simulations of 
the PDE (\ref{aa}) starting with the initial condition (\ref{bd}), with the same 
$\eta_{0}$, $p_{0}$, $q_{0}$ and $\Phi_{0}$.  

Since the collective coordinates approximation can work only for small perturbations, we choose a small driving amplitude $a=0.05$ for 
$f(x)=a e^{i K x}$. We also take $K=0.1$, which means that the spatial 
period of the forcing $L = 2 \pi/K \gg 1$. We are 
interested in periodic solutions and therefore we set the damping parameter $\beta=0$.  
Damped oscillatory solutions were already considered in Ref. \cite{fgm}. 

\begin{figure}[ht!]
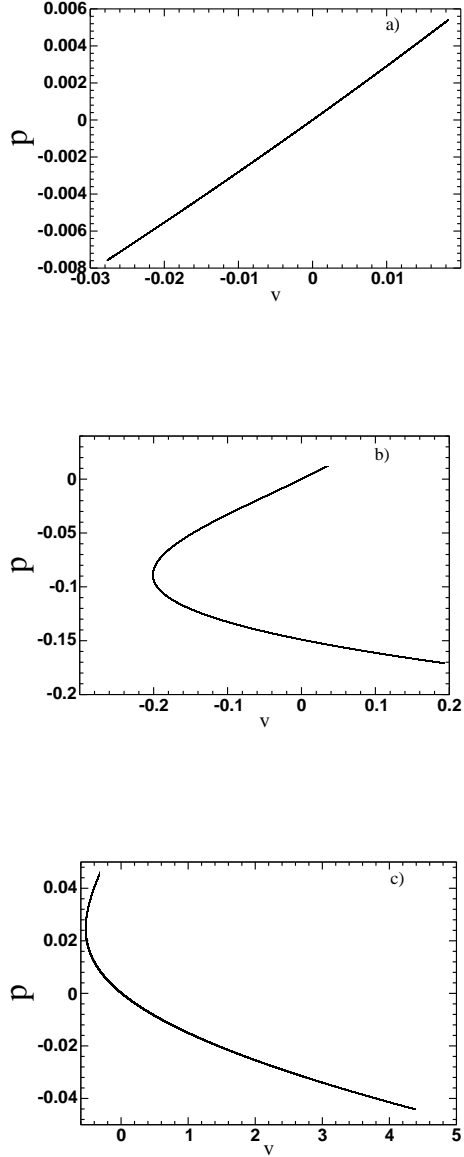

\begin{center}
\begin{tabular}{c}
\includegraphics[width=6.0cm]{fig1a.eps}  \\ 
\,\ \\
\\
\includegraphics[width=6.0cm]{fig1b.eps} \\
\,\ \\
\\
\includegraphics[width=6.0cm]{fig1c.eps}
\end{tabular}
\end{center}
\caption{Stability curve $p(\v)$ corresponding to $a=0.05$, $K=0.1$, 
$\delta=-1$ and $\beta=0$.  
The initial conditions for the Eqs.\ (\ref{be1})-(\ref{be4}) were 
$q_{0}=p_{0}=\Phi_{0}=0$ and a)  
$\eta_{0}=0.8$, b) $\eta_{0}=0.65$, c) $\eta_{0}=0.1$. 
The integration time $t_{f}=1000$.}
\label{fig1} 
\end{figure}

For $\delta \ge 0$, the $p(\v)$ curve predicts only unstable solitons 
which is confirmed by the simulations. 
For $\delta <0$, there are typically several stability regions which grow as 
$|\delta|$ is increased, while the parameters $a$ and 
$K$ are fixed \cite{fgm}. 
We concentrate here on the simplest case with only one 
stability and one instability region. Namely, we choose $\delta=-1$, 
$q_{0}=p_{0}=\Phi_{0}=0$ for 
which the soliton solutions are predicted to be stable if 
$\eta_{0} \ge \eta_{c}^{(1)}=0.684$ (Fig.\ \ref{fig1}a)  
and unstable for $\eta_{0} < \eta_{c}^{(1)}$. This is confirmed by 
our simulations of the PDE (\ref{aa}) to an accuracy of better than $1 \%$ 
in $\eta_{c}^{(1)}$. 

\vspace{1.0cm}

The range of initial amplitudes $\eta_{0}$ for which solutions of Eqs.\ 
(\ref{be1})-(\ref{be4}) 
feature a $p(\v)$ curve with a descending branch, can be divided into 
two subintervals, 
$0< \eta_{0} < \eta_{c}^{(2)}$ and $\eta_{c}^{(2)} < \eta_{0} < \eta_{c}^{(1)}$, 
where 
$\eta_{c}^{(2)}=0.288$. In the upper subinterval, $(\eta_{c}^{(2)}, \eta_{c}^{(1)})$, the curve 
$p(\v)$ exhibits two long branches, one with a positive and the other with a 
negative slope (Fig.\ \ref{fig1}b).  
In the lower subinterval, the positive-slope branch is short and very steep 
(Fig.\ \ref{fig1}c). The difference in the shape of the stability curves suggests 
different types 
of instability in the two subregions; however, in which way the 
instabilities are different cannot be deduced from the 
$p(\v)$ curve alone. 
To gain further insight into this difference, we plot a phase portrait 
for the dynamical system (\ref{be1})-(\ref{be4}). The vertical and horizontal 
axes in the 
portrait are chosen so that they admit a clear interpretation in terms of the full PDE, 
Eq.\ (\ref{aa}). To this end, we first 
transform to the frame of reference moving with the velocity $V_{f}$:  
\begin{equation}\label{new}
 u(x,t)=\Psi(X,t) e^{iKx}, \qquad X=x-V_{f} t,
\end{equation}
where $V_{f}=2 K$. Eq.\ (\ref{aa}) is taken to be an NLS driven by a 
space-time independent external force:   
\begin{equation}\label{ca}
 i \Psi_{t} + \Psi_{XX} + 2 |\Psi|^2 \Psi + (\delta - K^2) \Psi =a.
\end{equation}
(This equation was previously studied in a different context 
 \cite{igor3,Bar1,Bar2,Bar3} and two static soliton solutions were obtained explicitly \cite{igor3}. 
Unlike \cite{igor3,Bar1,Bar2,Bar3}, we focus here 
on moving solitons of Eq. (\ref{ca}).) 

Under the transformation (\ref{new}), the collective coordinates ansatz (\ref{bd}) becomes   
\begin{equation}\label{cb}
\Psi(X,t) = 2 i \eta \, \mathrm{sech}[2 \eta (X+V_{f} t -q)] 
e^{-i [(K-p) (X+V_{f} t)+p q + \Phi]}.
\end{equation}
The $\eta(t)$ and $p(t)$ components of the oscillatory solutions of (\ref{be1})-(\ref{be4})  
are periodic with period $T$, whereas $q$ and $\Phi$ are of the form  
 $q(t)= {\overline \v} t+ q^{(p)}(t)$, $\Phi(t)= -\alpha t + \Phi^{(p)}(t)$,  
where $q^{(p)}(t)$ and $\Phi^{(p)}(t)$ are $T$-periodic functions and $\alpha$ is a constant 
\cite{fgm}. The corresponding soliton (\ref{bd}), (\ref{cb}) has the mean velocity 
 ${\overline \v}$ in the original frame of reference, and ${\overline \v}-V_{f}$ in 
the moving frame. 

At the point  $x={\overline \v} t$ 
[or, equivalently, at $X= ({\overline \v}-V_f)t$], 
the function (\ref{cb}) has the following time dependence: 
\begin{equation}\label{cc}
\Psi = 2 i \eta \, \mathrm{sech}[2 \eta ({\overline \v} t -q)] 
e^{-i [K {\overline \v} t -p ({\overline \v} t  - q) + \Phi]}.
\end{equation}
The function (\ref{cc}) is a collective coordinates counterpart of the $\Psi$ field at the centre of the 
soliton solution of Eq.\ (\ref{ca}). Comparing Eq.\ (\ref{cc}) to the 
function $\Psi(X,t)|_{X= {\overline \v} t - V_{f} t}$ obtained in the 
direct numerical simulations of the full PDE (\ref{aa}), one can assess the validity and 
accuracy 
of the collective coordinates approximation. For this reason, we choose the complex function (\ref{cc}) as a 
representative of the four-dimensional dynamics, and plot its real versus 
imaginary part to 
generate the corresponding phase portrait. The soliton dynamics is described by 
the resulting orbits of the phase portrait. 

One can readily verify that these orbits are closed. Indeed, the modulus of the function 
(\ref{cc}) is periodic with period $T$. Therefore, to demonstrate the closure, 
one just needs to show that the argument of $\Psi$ changes by an integer multiple of 
$2 \pi$ over the period. We have $\arg \Psi = (\alpha -K {\overline \v}) t - 
p^{(p)} q^{(p)} - \Phi^{(p)}$, where the last two terms are 
$T$-periodic. As for the first term, the constants ${\overline \v}$ and $\alpha$ 
are found as the coefficients of the linearly-growing components of $q(t)$ 
and $\Phi(t)$, respectively. The numerical solution of Eqs.\ 
(\ref{be1})-(\ref{be4}) verifies that 
$\alpha -K {\overline \v}= 2 \pi/T$ in the stability range 
$\eta_{0} \ge \eta_{c}^{(1)}$, that  $\alpha -K {\overline \v}= -2 \pi/T$ 
in the lower instability subinterval $0 < \eta_{0} < \eta_{c}^{(2)}$, 
and $K {\overline \v}-\alpha=0$ in the 
upper instability subinterval $\eta_{c}^{(2)} < \eta_{0} < \eta_{c}^{(1)}$. These 
relations between ${\overline \v}$ and $\alpha$ hold to a  
numerical accuracy of $O(10^{-5})$.

\begin{figure}[ht!]
\begin{center}
\begin{tabular}{cc}
\psfrag{a}{$Re \Psi$}
\psfrag{B}{$Im \Psi$}
\includegraphics[width=8.0cm]{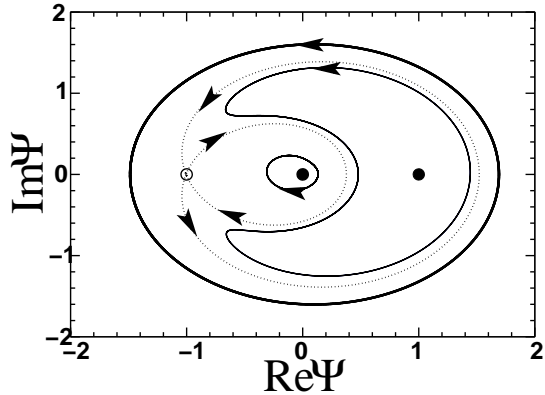}
\end{tabular}
\end{center}
\caption{The phase portrait of the system 
(\ref{be1})-(\ref{be4}) with $a$, $K$, 
$\delta$ and $\beta$ as in Fig.\ (\ref{fig1}). Shown is 
Im$ \Psi(X= {\overline \v} t-V_f t,t)$ versus Re$\Psi(X= {\overline \v} t-V_f t,t)$. 
 The large ellipse corresponds to $\eta_{0}=0.8$, the horseshoe to $\eta_{0}=0.65$ 
and the small ellipse to $\eta_{0}=0.1$. Other initial conditions are as in Fig.\ 
(\ref{fig1}). 
The separatrix  is shown by the dotted curve. 
The  filled and open circles are stable and unstable fixed points, 
respectively.  
}
\label{fig2} 
\end{figure}

Trajectories resulting from initial conditions in the interval 
$\eta_{0} > \eta_{c}^{(1)}$ are 
ellipses, with a positive sense of rotation (Fig. \ref{fig2}). 
The ellipses enclose a stable and an unstable fixed point on 
the real axis at about $+1$ and $-1$, respectively. 
(For the definition and calculation of these points see Appendix A). 
Fig. \ref{fig3}a compares the soliton amplitude $\eta(t)$ from collective coordinates theory 
to the amplitude measured in the direct simulations of Eq.\ (\ref{aa}).  

\vspace*{0.25cm}

For the upper instability subinterval 
$\eta_{c}^{(2)}<\eta_{0}<\eta_{c}^{(1)}$ the phase trajectory is a 
horseshoe 
(Fig.\ \ref{fig2}). 
This curve consists of an outer part with a positive sense of rotation and an inner part with a 
\emph{negative} sense of rotation relative to the origin. 
The two parts are correlated with the two branches with positive and negative 
slopes, respectively, of the $p(\v)$-curve in Fig.\ \ref{fig1}b. The soliton 
instability is seen in the simulation 
result in Fig.\ \ref{fig3}b. Note that the first harmonic vanishes after 
about $30$ time units, while 
the second harmonic persists. Eventually the soliton decays: the amplitude 
approaches zero  while the width tends to infinity.

\vspace*{0.5cm}

\begin{figure}[ht!]
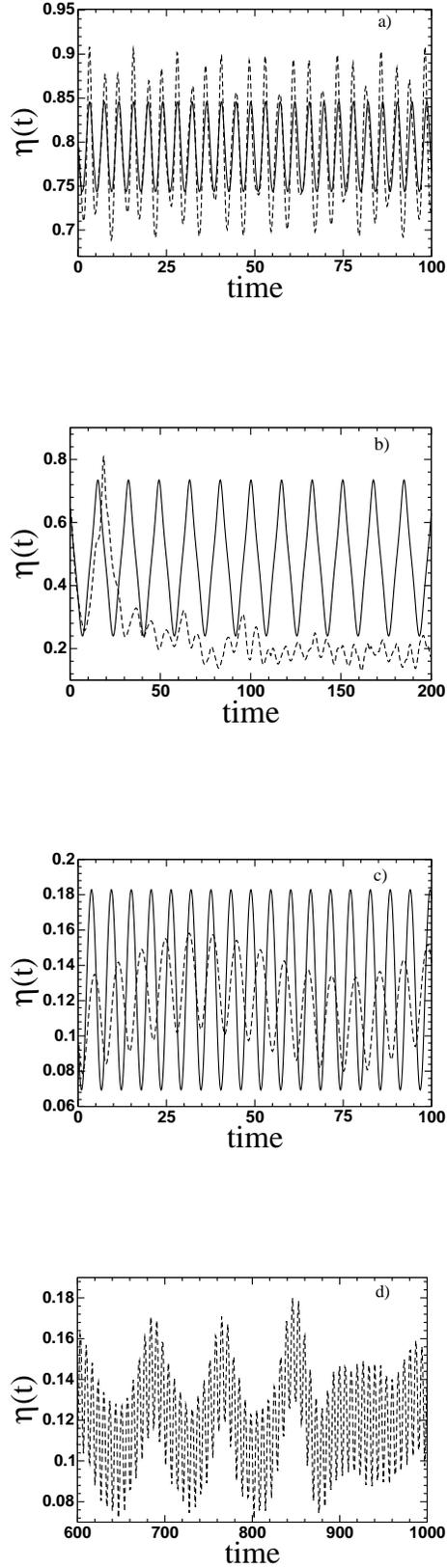

\begin{center}
\begin{tabular}{c}
\includegraphics[width=6.0cm]{fig3a.eps} \\
  \,\ \\
\\
\includegraphics[width=6.0cm]{fig3b.eps} \\
\,\ \\
\\
\includegraphics[width=6.0cm]{fig3c.eps} \\
\,\ \\
\\
\includegraphics[width=6.0cm]{fig3d.eps}
\end{tabular}
\end{center}
\caption{Soliton amplitude $\eta(t)$ from the collective coordinates theory (solid lines) and from the 
simulations (dashed lines). The parameters and the initial conditions are the   
same as in Fig.\ \ref{fig1}.  
a) $\eta_{0}=0.8$, b) $\eta_{0}=0.65$, c) $\eta_{0}=0.1$ 
(shown are results for early times $0 \le t \le 100$);  
d) $\eta_{0}=0.1$ (shown are simulation results for late times $600 \le t \le 1000$.) 
}
\label{fig3} 
\end{figure}

\vspace*{0.5cm}

\begin{figure}[ht!]
\begin{center}
\begin{tabular}{c}
\includegraphics[width=8.0cm]{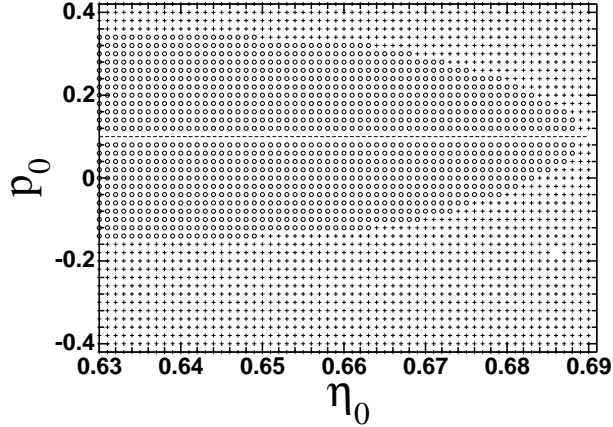}
\end{tabular}
\end{center}
\caption{Stability diagram near $\eta_{c}^{(1)} = 0.684$,  
with $q_{0} = \Phi_{0} = 0$. Parameters: $a = 0.05$, $K =0.1$, 
$\delta = -1$, $\beta = 0$. Circles: unstable soliton. Plus: stable 
soliton. Dashed line: for $p_0=K$, the $p(\v)$ curve is 
a point.}
\label{fig3a} 
\end{figure}

For the lower instability interval $0 < \eta_{0} < \eta_{c}^{(2)}$ the situation is quite different, both in the collective coordinates theory and in the simulations: 
The phase portrait features an ellipse, but with 
the \emph{negative} sense of rotation (Fig.\ \ref{fig2}). Moreover, 
the ellipse is much smaller than the one arising in the 
stability region so that it encloses only one fixed point. 
This fits with the simulations in which the soliton remains metastable 
for a relatively long time, exhibiting a periodic modulation 
of the oscillation amplitude  
(Fig.\ \ref{fig3}c), but then the instability sets in (Fig.\ \ref{fig3}d).

So far we have varied $\eta_0$, with $p_0 = q_0 = \Phi_0 = 0$ 
kept fixed. We now consider the stability diagram in the $\eta_0$-$p_0$ 
plane near the critical value $\eta_{c}^{(1)}$, which separates the stability 
interval from the upper instability interval (above). 
Fig. \ref{fig3a} shows that a finite value of the normalized momentum $p_0$ 
stabilizes the soliton and therefore the stability region is enlarged. 
The curve which separates stability and instability regions is roughly 
a parabola.

Finally we would like to emphasize the crucial importance of using the 
\emph{normalized} momentum $p$ in the stability analysis ---rather than the canonical momentum 
$P$ (as was proposed in \cite{fgm}). We have established that the empirical 
stability criterion suggested in \cite{fgm} disagrees with the 
results of numerical simulations when the initial normalized momentum 
is too large, i.e. $p_0 > K$  (for positive $K$).  
Let, for instance, the parameters of the equation take the same values 
as in Fig.\ \ref{fig1} ($a=0.05$, $K=0.1$, $\delta=-1$ and $\beta=0$), 
and take the same initial conditions as for the stable stationary 
solution in the Appendix ($\eta_0= 0.5 \sqrt{K^2-\delta}$, $\Phi_0=\pi/2$, 
$q_0=0$), except that  this time $p_0=K+d$, where $0<d<0.2$. 
In this case the numerical solutions of the collective coordinates equations can be represented 
in a very good approximation by $p(t)=p_0 + a_p (1-\cos \Omega t)$ 
and   $\v(t)=\v_0 + a_\v (1-\cos \Omega t)$, where $p_0$, $\v_0$, $a_p$, $a_\v>0$. 
Thus $p(\v)$ is a straight line with slope $a_p/a_\v >0$. 
This predicts stability, 
the same 
as the orbit in the phase portrait which is a small ellipse with positive 
sense of rotation around the stable fixed point at about $+1$ on  
the real axis. The stability is 
confirmed by simulations. However, when the momentum $P=4 \eta p$ is 
used the situation is different. $\eta(t)$ can be expressed via $p(t)$ 
by using the exact relation $\eta=\eta_0 (p_0-K)/(p-K)$ ($p \ne K$), which is obtained 
from Eqs.\ (\ref{be1}) and (\ref{be3}) where one integration has been carried out. 
Finally, $p(\v)$ from above is inserted and one can see that $P$ decreases when 
$\v$ increases and viceversa. Thus the slope $dP/d\v <0$ 
predicts instability which disagrees 
with the simulations.

\section{Harmonic $K(t)$} \label{sec4}

As stated in the Introduction, one of the aims of this paper is to 
verify whether the stability criterion $p'(\v)>0$ remains 
applicable to time-dependent forces of the form 
$f(x,t)= a e^{i K(t) x}$.  
In this section we consider the case of a harmonically modulated forcing 
wavenumber: 
\begin{eqnarray}\label{eq4.1}
K(t) & = & k \sin(\omega t + \theta),
\end{eqnarray}
first without a damping term in the NLS ($\beta=0$), 
then with the damping ($\beta > 0$). 

We choose the same parameters as in Section \ref{sec3}: $a=0.05$, $k=-0.1$ 
which implies  
$|K| \ll 1$. In order to be in the adiabatic regime we choose 
a small modulation frequency 
$\omega=0.02$. Finally, we let $\theta=0$ and choose initial conditions 
$q_{0}=p_{0}=0$, $\eta_0=1$, and $\Phi_{0}=\pi/2$. 

\begin{figure}[ht!]
\begin{center}
\begin{tabular}{c}
\includegraphics[width=7.0cm]{fig5a.eps}  \\ 
 \,\ \\
\\
\includegraphics[width=7.0cm]{fig5b.eps}  \\
 \,\ \\
\\
\includegraphics[width=7.0cm]{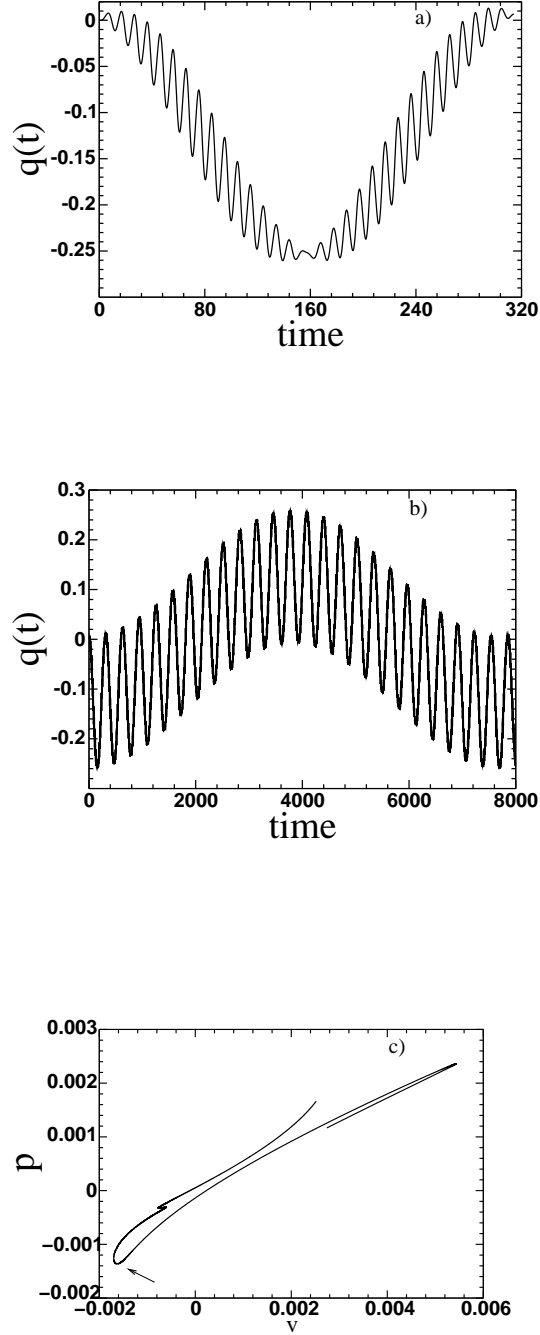}
\end{tabular}
\end{center}
\caption{
Collective coordinates results for harmonic $K(t)$, no damping. $a=0.05$, $k=-0.1$, $\omega=0.02$, $\theta=0$, $\delta=-3$, $\beta=0$, 
$q_{0}=p_{0}=0$, $\Phi_{0}=\pi/2$, $\eta_{0}=1$. 
a) $q(t)$  exhibits  
$\omega_{i}$-oscillations modulated by the frequency $\omega$.  
Shown is the interval $0 \le t \le T_{d}= 2 \pi/ \omega$.  
b) $q(t)$  exhibits 
$\omega$-oscillations modulated by the frequency $\omega_{l}=2 \pi/T_{l}$. 
Here $0 \le t \le 8000$. 
c) Stability curve $p(\v)$ for $T_{d}/2 -5 \le t \le T_{d}/2 +10$. 
The arrow points to the section of the curve with a 
negative slope loop.}
\label{fig5} 
\end{figure}

The numerical solutions of the collective coordinates equations 
(\ref{be1})-(\ref{be4}) exhibit oscillations with three very different 
frequencies in their spectrum. This is most explicit in the behaviour of 
$q(t)$ (Fig.\ \ref{fig5} a,b). First, there are intrinsic oscillations with the frequency 
$\omega_{i}$; these have a period $T_{i}$ of the order of $10$, similarly to the oscillations in the case of 
the constant $K$ discussed in the previous section. Second, there are oscillations 
with the driving frequency $\omega$ whose period 
$T_{d}=2 \pi/ \omega \approx 314$. 
Finally, there are oscillations with a very low frequency 
$\omega_{l}$ and  very long period $T_{l} \approx 8000$ (Fig.\ \ref{fig5}b). 
The resulting 
stability curve $p(\v)$ exhibits many small loops which have a short 
section with a negative slope. An example is 
given in Fig.\ \ref{fig5}c. (For clarity the curve is plotted only 
over a short time interval). 
The negative slope predicts  instability; this is confirmed by our simulations of 
the full PDE.

\vspace{0.5cm}

\begin{figure}[ht!]
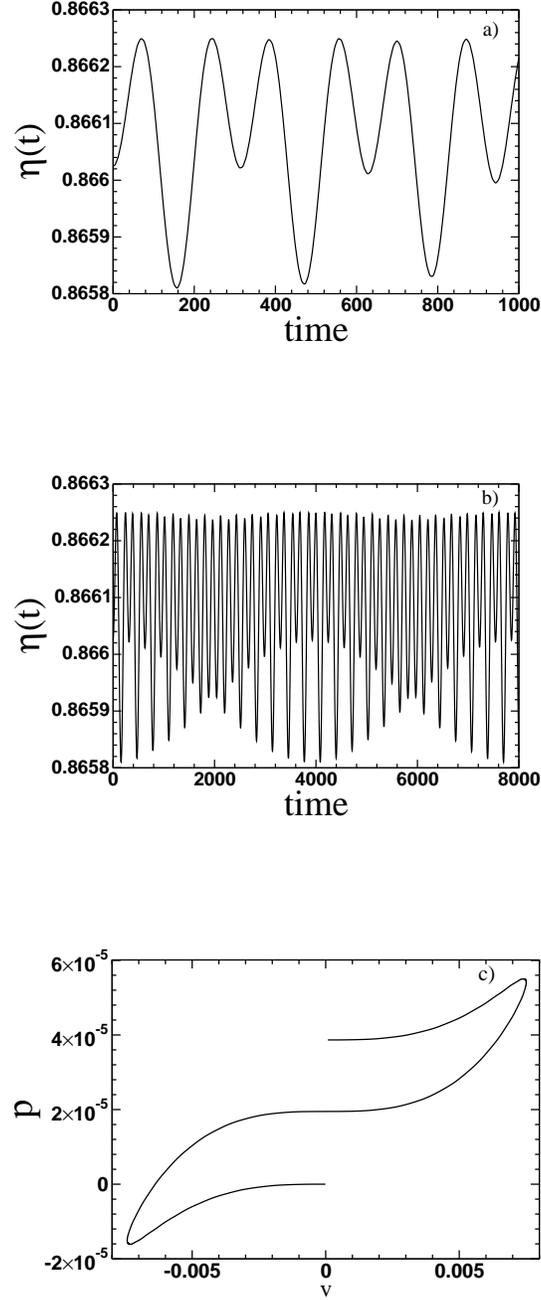

\begin{center}
\begin{tabular}{cc}
\includegraphics[width=7.0cm]{fig6a.eps} \\
 \,\ \\
\\
\includegraphics[width=7.0cm]{fig6b.eps} \\
 \,\ \\
\\
\includegraphics[width=7.0cm]{fig6c.eps} \\ 
\end{tabular}
\end{center}
\caption{Collective coordinates results for harmonic $K(t)$, no damping. 
Same parameters and initial conditions as in Fig.\ \ref{fig5}, but 
$\eta_{0}=\sqrt{-\delta}/2$. (For this choice the intrinsic $\omega_{i}$-oscillations 
vanish.)   
a) The amplitude $\eta(t)$ exhibits no $\omega_{i}$-oscillations, 
only $\omega$-oscillations 
modulated by $\omega_{l}$-oscillations which are hardly visible for 
$0 \le t \le 1000$. 
b) For $\eta(t)$ for $0 \le t \le 8000$ both $\omega$- and $\omega_{l}$-oscillations 
are visible.
c) Stability curve for $0 \le t \le T_{d}$.}
\label{fig6} 
\end{figure}

Stable solitons can be obtained by changing $\eta_{0}$ in such a way that the 
loops do not arise. This is achieved by suppressing the intrinsic oscillations, 
since their period $T_{i} \approx 10$ is of the same order as the time scale of the loops, see Fig.\ 
\ref{fig5}c. The intrinsic oscillations disappear when we choose $\eta_{0}=\sqrt{-\delta}/2$ 
(Fig.\ \ref{fig6}a,b). In this case $\eta(t)$ performs very small oscillations around $\eta_{0}$ 
and the two dominant terms on the r.h.s. of 
Eq.\ (\ref{be4}), namely $-4\eta_{0}^2$ and $-\delta$, cancel each other. Fig.\ \ref{fig6}c 
demonstrates that the small loops have indeed disappeared. 

The resolution of Fig.\ \ref{fig6}c does not allow verification of whether 
there are sections with negative slope near the turning 
points of the stability curve. We now show that there cannot 
be any, as the curve develops cusps at the turning points. 
Consider the region around one of the maxima (or minima) 
of the $\omega$-oscillations of the collective coordinates (Fig.\ \ref{fig6}a). 
The functions $q(t)$, $\eta(t)$ 
etc. are \emph{not} symmetric with respect to $t_{m}$ (position of the extremum) 
due to the existence of the very slow $\omega_{l}$-oscillations. 
The same holds for $p$; hence 
\begin{eqnarray} \label{eq26}
p(t)= \left \{
\begin{tabular}{ll}
$p_{m}-C_{l} (t-t_{m})^2$ & \quad for $t \le t_{m}$ \\
$p_{m}-C_{r} (t-t_{m})^2$ & \quad for $t \ge t_{m}$
\end{tabular}
\right. 
\end{eqnarray}
with $C_{l} \ne C_{r}$. For the velocity $\v(t)=\dot{q}(t)$ the asymmetry is 
negligible, compared to the asymmetry of $p(t)$, because the time derivative 
$\dot{q}$ contains a factor $\omega_{l} \ll 1$. Thus 
$\v(t)=\v_{m}-b (t-t_{m})^2$ for both $t \le t_{m}$ and $t \ge t_{m}$. 
Eliminating $t$ we obtain $p=p_{m}-C_{l} (\v_{m}-\v)/b$ as $\v$ increases up to 
its maximal value $\v_{m}$, and $\v=\v_{m}-C_{r} (\v_{m}-\v)/b$ as $\v$ decreases 
from $\v_{m}$. Thus the stability curve $p(\v)$ has a cusp with two different 
(but positive) slopes $C_{l}/b$ and $C_{r}/b$ at the turning point $\v=\v_{m}$. 
The absence of segments of the curve with $dp/d\v<0$  
predicts stability for the soliton. This is confirmed by the simulations of the 
PDE (\ref{aa}). 

\vspace{1.0cm}

\begin{figure}[ht!]
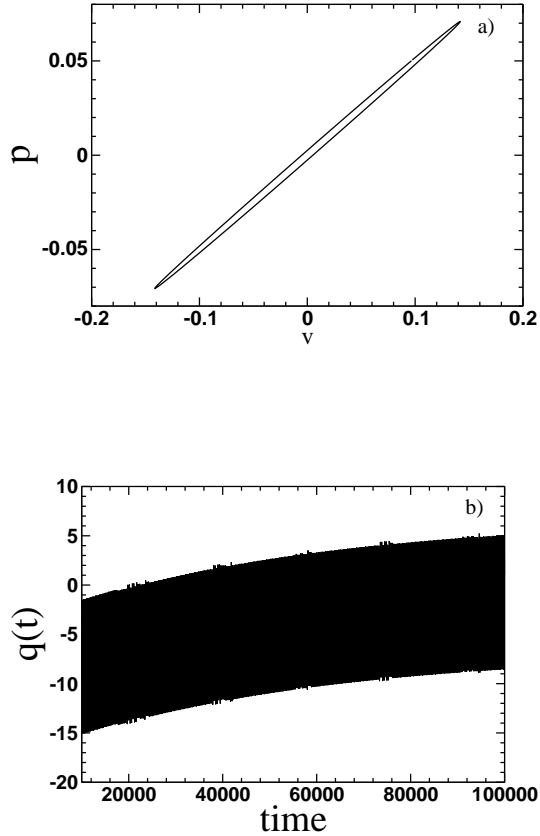

\begin{center}
\begin{tabular}{cc}
\includegraphics[width=7.0cm]{fig7a.eps} \\
 \,\ \\
\\
\includegraphics[width=7.0cm]{fig7b.eps} 
\end{tabular}
\end{center}
\caption{Collective coordinates results for harmonic $K(t)$, 
nonzero damping. Same data as in Fig.\ \ref{fig6}, but 
$\eta_{0}= 0.866 \approx \sqrt{-\delta}/2$, $\beta=0.01$. 
a) stability curve $p(\v)$ for $3 T_{d} \le t \le 4 T_{d}$. 
b) $q(t)$ for $10 000 \le t \le 100 000$.}
\label{fig7} 
\end{figure}

When the damping term $-i \beta u$ is included in the r.h.s. of the 
NLS equation (\ref{aa}), the collective coordinates dynamics simplifies. Namely, both the 
intrinsic oscillations and the low-frequency oscillations are damped 
out from solutions of the collective coordinates equations after a transient time 
$t_{tr}=1/\beta$. After this transient, all collective coordinates oscillations become locked to the 
driving frequency $\omega$. 
The stability curve in this case consists of two nearly-straight lines which form 
sharp cusps at both ends 
(Fig.\ \ref{fig7}a). Thus there are no sections with a negative slope and the soliton is 
predicted to be stable. This is confirmed by the simulations of the PDE. For long times 
($t \gg t_{tr}$) the average soliton velocity 
${\overline \v}$ slowly approaches zero  
(Fig.\ \ref{fig7}b); this behaviour is independent of the initial conditions. 
Thus there is no unidirectional motion of the soliton for long 
times; the reason will be established in the next section. 

As $\beta$ is decreased, the stability curve becomes wider and the decay 
of ${\overline \v}$ to zero faster. On the contrary, as 
$\beta$ is increased, the stability curve 
becomes narrower, 
and ${\overline \v}$ decreases to zero more slowly. However, for $\beta$ above a 
critical value $\beta_{c}$ 
($\beta_{c}\approx0.035$ for the parameters of Fig.\ \ref{fig7}) the collective coordinates solutions become 
unstable. Direct simulations also confirm the soliton's instability.    

\vspace{0.5cm}

\section{Biharmonic driving: ratchets} \label{sec5}
The simplest ratchet models consider a point-like particle in a 
periodic potential driven by an AC force $f(t)$. Under certain conditions related to the breaking of symmetries, unidirectional motion of the particle can take place 
despite the applied force having zero temporal average 
\cite{hanggi-bartu,physicstoday,brownian,mm,rei}. Particle 
ratchets were generalized to nonlinear field theoretic systems, in which 
particles are replaced by solitons 
\cite{linke,zapata,marche,exp,exp1,salerno2,cos}. In particular, solitons  
in nonlinear Klein-Gordon systems can move on the average in one direction, if either a 
temporal or a spatial symmetry is broken. 

A temporal symmetry, namely a time-shift symmetry, is broken by a 
biharmonic force  
\cite{gorbach0,flach}. In this case the mechanism of the ratchet effect 
was clarified by a collective coordinates theory employing the soliton position and width as 
collective coordinates  
\cite{sazo,prl2003,chaos}. Due to the coupling between the translational and 
internal degrees of freedom, energy is pumped nonuniformly into the system, generating 
a unidirectional motion. The breaking of the time-shift symmetry gives rise to a 
resonance mechanism that is present whenever the soliton 
oscillation spectrum comprises at least one of the frequency components 
of the driving force. 

In this section we investigate whether the NLS solitons show a behavior similar 
to the Klein-Gordon kinks.  
This is particularly interesting since the NLS solitons are non-topological, 
whereas 
the vast majority of reports on soliton ratchets have so far focussed on 
topological solitons. 

We consider the NLS (\ref{aa}) where the perturbation
\begin{eqnarray}\label{eq27}
R = f(x,t) - i \beta u,
\end{eqnarray}   
has the form 
\begin{eqnarray}\label{eq28}
f(x,t) = a_{1} e^{i K_{1}(t) x} + a_{2} e^{i K_{2}(t) x}, 
\end{eqnarray}
with
\begin{eqnarray}\label{eq29}
K_{1} = k_{1} \sin(\omega t), \qquad K_{2} = k_{2} \sin(2 \omega t+\theta). 
\end{eqnarray}
Consider first the single-harmonic case: $a_{2}=0$. When  
$t \gg t_{tr}= 1/\beta$, the soliton oscillations are locked to the driving 
frequency 
$\omega$ and are independent of the initial conditions (see Section IV). 
Thus there exists  
a global solitonic attractor. 

We now perform a symmetry analysis 
\cite{gorbach,flach}. The perturbed NLS is invariant under the symmetry operation 
\begin{eqnarray}\label{eq30}
\mathcal{S}\,:\,t\mapsto t + T/2, \quad x \mapsto -x.
\end{eqnarray}
At the same time, the transformation $\mathcal{S}$ changes the sign of the 
soliton velocity 
$\v(t)=\dot{X}(t)$. The soliton position is defined by 
\begin{eqnarray}\label{eq31}
X(t)= \frac{\int_{-\infty}^{+\infty} dx\, x \rho(x,t)}
{\int_{-\infty}^{+\infty} dx\, \rho(x,t)}
\end{eqnarray}
with 
\begin{eqnarray}\label{eq32}
\rho(x,t) = ||u(x,t)|^{2}-|u_{bg}(x,t)|^2|.
\end{eqnarray}
Here $u_{bg}(x,t)=a_{bg}(t) \exp(i K_{1} x)$ 
is the background field  
to which the soliton decays as 
$|x| \to \infty$ \cite{gorbach,fgm}. When $|u(x,t)|^2$ from the simulations 
is plotted, the soliton sits on a shelf with homogeneous intensity 
$|a_{bg}(t)|^2$.  
The shelf has little influence on the soliton dynamics \cite{fgm}; 
this is why the collective coordinates theory is in a good agreement with simulations, despite ignoring  
the presence of the background. 

Since 
the attractor is global, the transformation $\mathcal{S}$ maps it onto itself. 
This implies that the average velocity on the attractor is zero. The soliton performs 
periodic oscillations about its equilibrium position which are reproduced 
by the collective coordinates theory 
(Fig.\ \ref{fig7}b). 

In order to construct a solitonic ratchet, i.e. obtain a stable soliton with 
${\overline \v} \ne 0$, it is necessary to break the 
invariance under the operation $\mathcal{S}$. The simplest 
way to do this is to employ the \emph{biharmonic} driving in Eq.\ (\ref{eq28}) with 
$a_{1} \ne 0$ and $a_{2} \ne 0$. 
The collective coordinates equations (\ref{be1})-(\ref{be4}) can easily be extended to the case of the forcing 
function $f$ including two terms. 
In particular, Eq.\ (\ref{be1}) is replaced with 
\begin{eqnarray}
\label{be1a}
\dot{\eta} &=& - 2 \beta \eta - \sum_{i=1}^{2} a_{i}  \frac{\pi}{2}\,   \mathrm{sech} 
A_{i} \cos B_{i},
\end{eqnarray}
where
\begin{eqnarray}
\label{bf1a}
A_{i} &=& \frac{\pi}{4 \eta(t)} [K_{i}(t) -p(t)], \\
\label{bf2a}
B_{i} &=& \Phi(t) + K_{i}(t) q (t),
\end{eqnarray}
while the $K_{i}$ are as in Eqs. (\ref{eq29}). The collective coordinates 
equations for $\dot{q}$, $\dot{p}$ and $\dot{\Phi}$ are modified in a similar 
way. 

Since the collective coordinates description is accurate only for small perturbations, we take small driving 
amplitudes $a_{1}=a_{2}=0.05$. We choose a very small driving frequency $\omega=0.002$ 
in order to remain in the adiabatic regime. If the damping coefficient $\beta$ is 
chosen too large, the soliton amplitude $\eta$ quickly relaxes to zero while  
$q(t)$ and $\Phi(t)$ rapidly go to 
infinity.  
For example, for the parameters $\delta=-3$, $k_{1}=k_{2}=k=0.001$, $\theta=0$ and the 
IC $\eta_{0}=1$, $q_{0}=p_{0}=0$, 
$\Phi_{0}=\pi/2$, this instability occurs when $\beta > 0.065$. 
On the other hand, 
if $\beta$ is chosen too small (e.g., $\beta=0.01$), the average 
soliton velocity grows without bound over sufficiently long integration times  
($t_{f} \sim 10^{5}$).  
Thus we can expect a 
stable ratchet effect only for intermediate values of $\beta$, for instance 
$\beta=0.04$. As we find that ${\overline \v} \sim k$, a larger ratchet effect can be obtained by 
increasing $k$. However, when $k$ exceeds a certain critical value $k_{c}$, the average 
velocity starts to grow slowly with time. (For the chosen parameter values, 
$k_{c}=0.002$).  

Using the parameter values and initial conditions for which the collective coordinates equations exhibit 
stable solutions we perform direct simulations of Eq.\ (\ref{aa}). 
Our aim is to test whether an initial waveform  (\ref{bd}) will evolve into 
a stable solitary wave over the time  $t_{tr}=1/\beta$. However, it turns 
out that the initial structure 
evolves rapidly immediately after the start of the simulation and quickly 
decays to zero.  

In order to obtain stable solitary waves we need to 
improve the initial conditions. 
This can be achieved by setting the initial conditions 
equal to the mean values about which the collective coordinates oscillate, 
once the transients have elapsed. 
These mean values can be obtained from an approximate analytical solution of the collective coordinates equations:
We let  
\begin{eqnarray}\label{eq36}
q &=& {\overline \v} t + C_{q}, \\  \nonumber
p &=& \bar{p} + C_{\xi}, \\ \nonumber 
\eta &=& \bar{\eta} + C_{\eta}, \\ \nonumber
\Phi &=& \bar{\Phi} + C_{\Phi}, 
\end{eqnarray}    
where $C_{x}$ are oscillations with amplitude $a_{x}$ and zero mean. 
We choose $\omega=O(10^{-3})$, 
and $k_{1}=k_{2}=k=O(10^{-3})$. The other parameters 
$(\delta,a_{1},a_{2},\beta)$ do not have to be small for the 
following perturbation analysis and can therefore be chosen in $O(1)$. 
Substituting in the collective coordinates equations we retain 
only the leading terms in the perturbation series. This gives  
\begin{eqnarray} \label{eq37}
\bar{p} &=&0, \quad p=\frac{a_{1} K_{1}(t)+a_{2} K_{2}(t)}{(a_{1}+a_{2})}=O(10^{-3}), \\ \label{eq38}
\bar{\eta} &=& \frac{1}{2} \sqrt{-\delta}, \quad a_{\eta}=O(10^{-6}), \\ \label{eq39}
\bar{\Phi} &=& \arccos\left(\frac{-4 \beta \bar{\eta}}{\pi(a_{1}+a_{2})}\right), \quad a_{\Phi} = O(10^{-3}), \\ 
\label{eq40}
q &=& {\overline \v} t + \frac{k}{\omega} \frac{2 a_{1}}{a_{1}+
a_{2}} (1-\cos(\omega t))+ \frac{k}{2 \omega} 
\frac{2 a_{2}}{a_{1}+a_{2}}\\ 
& \, &  \times  (1-\cos(2 \omega t+\theta)), \quad {\overline \v}=O(10^{-6}).
\end{eqnarray} 
Eqs.\ (\ref{eq37})-(\ref{eq40}) are in very good agreement with 
the numerical solution of the collective coordinates equations. 
We note that the constants $\bar{p}$, $\bar{\eta}$ and $\bar{\Phi}$ do not depend 
on the relative phase $\theta$, but 
the variable components  of $q(t)$ and $p(t)$ do.  

\begin{figure}[ht!]
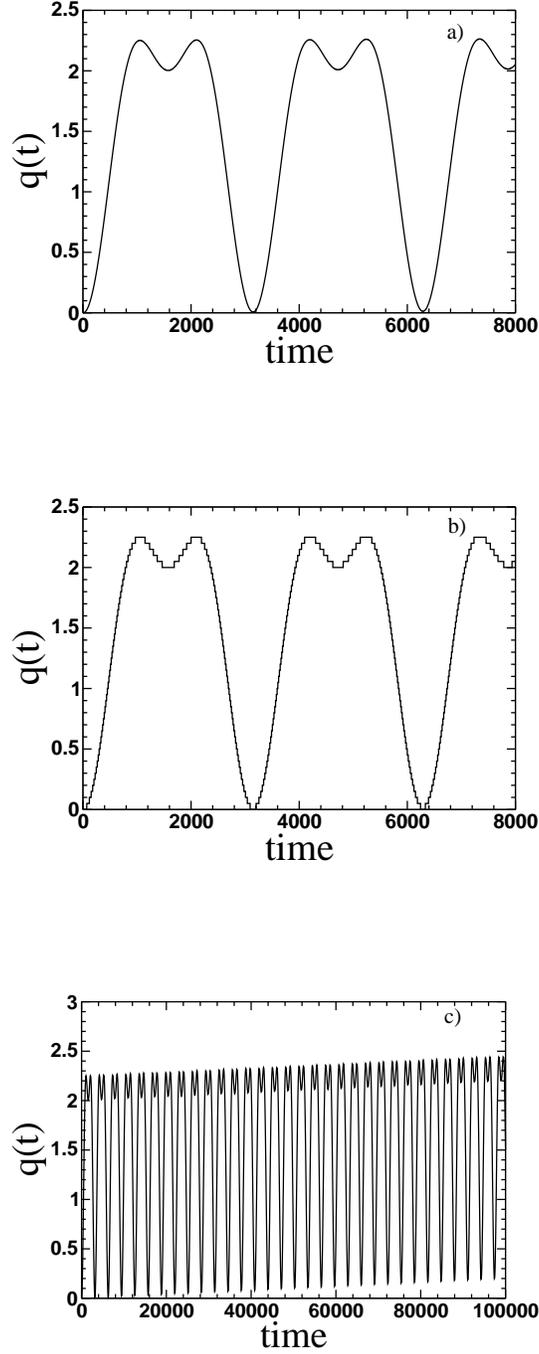

\begin{center}
\begin{tabular}{c}
\includegraphics[width=7.0cm]{fig8a.eps}  \\
\,\ \\
\\
\includegraphics[width=7.0cm]{fig8b.eps}  \\
\,\ \\
\\
\includegraphics[width=7.0cm]{fig8c.eps}  \\
\end{tabular}
\end{center}
\caption{
Soliton position $q(t)$: 
(a) from the collective coordinates theory,  
(b) from simulations,  
(c) the ratchet effect in the collective coordinates theory visible over long times ($t_{f} \approx 30$ periods). 
Parameters: $a_{1}=a_{2}=a=0.05$, $k_{1}=k_{2}=k=0.002$, 
$\delta=-3$, $\omega=0.002$, $\theta=0$,  $\beta=0.08$,  
with initial conditions $\eta_{0}=\bar{\eta}$, 
$q_{0}=p_{0}=0$, $\Phi_{0}=\bar{\Phi}$. 
}
\label{fig8} 
\end{figure}

The improved initial conditions are now: 
$p_{0}=\bar{p}=0$, $\eta_{0}=\bar{\eta}$, 
$\Phi_{0}=\bar{\Phi}$, $q_{0}=0$. After a transient time the 
numerical trajectory settles to the solution (\ref{eq37})-(\ref{eq40}). 
This yields $\v(t)=\dot{q}=2 p$ and thus 
$p(\v)=\frac{1}{2} \v$ is a straight line with  positive slope. 
 Our stability criterion predicts the stability of the 
soliton and simulations of the PDE 
confirm this (Fig. \ref{fig8}a,b). 
As ${\overline \v}$ is very small, the 
ratchet  effect is not visible on the time scale of Fig. \ref{fig8}a,b, but 
can be observed 
over longer simulation times $t_{f} \approx 30 T_{d}$ (Fig. \ref{fig8}c).
The soliton collective coordinates amplitude $\eta(t)$ oscillates about $0.866025$ in the interval 
$[0.866024, 0.866026]$, in agreement with Eq.\ (\ref{eq38}). 
In the simulations, $\eta(t)$ oscillates about  
$0.87363$ in $[0.87305, 0.87390]$. 

The ratchet effect is also observed for higher driving frequencies   
(e.g., $\omega=0.01$ and $0.02$). However, the average velocity 
${\overline \v}$ decreases in proportion to  
$1/\omega$, similar to the last two terms  
in Eq. (\ref{eq40})). 
It is important to emphasize here that we only succeeded in determining  
initial conditions 
for a stable soliton thanks to the 
availability of the explicit solution of the collective coordinates equations 
and our 
stability criterion. It would be very difficult  
to identify the corresponding small basin of attraction via numerical simulations of 
the PDE. 
 
\vspace{0.5cm}

Finally, we discuss the dependence of the average velocity ${\overline \v}$ on 
the relative phase $\theta$ in the biharmonic driving force  
(\ref{eq29}). As expected for a ratchet system with biharmonic driving 
\cite{flach,chaos}, ${\overline \v}(\theta)$ is sinusoidal with the period 
$2\pi$. It attains its maximum value near 
 $\theta=0$ and its small negative minimum value 
near $\theta=\pi$. The size and shape of the basin of attraction  
around $(\bar{\eta},\bar{\Phi})$ also depend strongly on $\theta$; this effect will be examined in a 
future work. 

\section{Summary}

We have formulated a refined empirical stability criterion for the driven NLS solitons. 
Unlike stability criteria available in the literature, the new criterion is based on a 
Collective Coordinate (CC) description. Solving  (analytically or numerically) 
evolution equations for 
the four collectives coordinates, we use the 
resulting trajectories to evaluate the normalized soliton momentum $p(t)$
and the soliton velocity $\v(t)$.  
These give a parametric ``\textit{stability curve}'', $p(\v)$. 

Whenever the curve $p(\v)$ has a section with a negative slope ($dp/d\v<0$), 
we observe the instability of the soliton in direct numerical simulations. 
We therefore conjecture that the availability of a section with a negative 
slope is a sufficient condition for the instability of the soliton. 
We do not have a 
mathematical 
proof of this conjectured criterion; however we have verified it in a variety of 
situations using constant, harmonic and biharmonic functions $K(t)$, with 
or without the damping term.

 Establishing a theoretical justification of this 
conjecture is a subject of future work,  
first for the cases of $K(t)$ which we have considered in this paper, 
then for a general function $K(t)$. 
One of the foreseen difficulties is related to  
the fact that the soliton solution of the driven NLS does not vanish 
as $x \to \pm \infty$, because the perturbation $f(x,t)=a \exp[i K(t) x]$ 
does not decay to zero 
in these limits. On the other hand,  
our collective coordinates theory is based on a soliton ansatz which vanishes as 
$x \to \pm \infty$.

    For the case of constant $K$ and zero damping, all collective 
coordinates perform periodic motions.  
This allowed us to compute a phase portrait which consists of closed orbits 
on a complex plane. 
The soliton evolution is described by motion along one of these 
orbits. 
We observe that the sense of rotation of the orbit is correlated with 
the stability/instability of the soliton (determined 
in simulations  of the full PDE). Namely,  
\begin{enumerate}
\item If the orbit is an ellipse with a positive sense of rotation, the soliton is stable. 
\item If the orbit is a horseshoe where the inner part has  
negative  and the outer part positive sense of rotation, the soliton 
is unstable and desintegrates very quickly. 
\item If the orbit is an ellipse with a negative sense of rotation, 
the soliton remains metastable for a 
relatively long time but eventually desintegrates.
\end{enumerate}

An interesting question is whether our collective coordinate 
approach and our stability criteria can also be applied to NLS equations 
with a more general form of the nonlinearity. Work is in progress regarding  
the case of a nonlinearity with arbitrary power, $(u^{*} u)^{\kappa}$, 
where $\kappa = 1$ corresponds to the NLS of this paper. 
The unperturbed NLS has stable solitons for $0 < \kappa <2$ and it will be 
interesting to determine how the stability of the solitons is affected by the 
perturbation $f(x,t)$. 

\section{Acknowledgments}
F.G.M. acknowledges the hospitality of the Mathematical Institute of the University of Seville (IMUS) 
and of the Theoretical Division and Center for Nonlinear Studies at the 
Los Alamos National Laboratory. Work at Los 
Alamos was supported by USDOE. F.G.M.  acknowledges financial supports by
the Plan Propio of the University of Seville and by Junta de Andaluc\'{\i}a 
under the grant IAC09-III-6399. 
N.R.Q. acknowledges financial support
by the DAAD under the grant A/08/04067, by the Ministerio de Educaci\'on y Ciencia (MEC, Spain)
through FIS2008-02380/FIS, and by Junta de Andaluc\'{\i}a
under the projects FQM207, FQM-00481,  P06-FQM-01735 and P09-FQM-4643.

\section{Appendix A: Fixed points of the phase portrait} \label{apA}

For the case of the time-independent force $f(x)=a e^{i K x}$ and zero damping, 
we adopt the 
following ansatz for stationary solutions of the collective coordinates equations:
\begin{equation}\label{A1}
 q(t)=\v_{s} t, \quad \eta(t)=\eta_{s}, \quad p(t)=p_{s}, \quad \Phi(t)=\Phi_{s}-\alpha_{s}t.
\end{equation}
Eq.\ (\ref{be1}) yields $\cos B\equiv 0$ which results in
\begin{equation}\label{A2}
 K \v_{s}=\alpha_{s}, \quad \Phi_{s}^{\pm} = \pm \frac{\pi}{2}, \quad \sin B=\pm 1.
\end{equation}
We insert the ansatz (\ref{A1}) in Eq.\ (\ref{cc}) yielding 
\begin{equation}\label{A5}
 \Psi_{s} = 2 i \eta_{s} \, \mathrm{sech}[2 \eta_{s} ({\overline \v}-\v_{s}) t]
{e}^{-i [-p_{s} ({\overline \v}-\v_{s}) t + \Phi_{s} + 
(K {\overline \v}-\alpha_{s}) t]}.
\end{equation}
The fixed points of the phase portrait correspond to the time-independent 
$\Psi$, 
i.e. ${\overline \v}=\v_{s}$. 
Using Eqs.\ (\ref{A2}), we obtain two fixed points
\begin{equation}\label{A6}
 \mathrm{Re} \Psi_{s}^{\pm} = \pm 2 \eta_{s}, \qquad \mathrm{Im} \Psi_{s}^{\pm} =0.
\end{equation}
The $\pm$ signs refer to the two cases in Eqs.\ (\ref{A2}).

Combining Eqs.\ (\ref{be2}) and (\ref{be4}) with Eq.\ (\ref{A2}), $\alpha_{s}$ can be eliminated 
and we are left with two equations 
\begin{eqnarray}
 \v_{s}&=&2 p_{s} \pm \frac{a \pi^2}{8 \eta_{s}^2}\,  \mathrm{sech} A_{s}  
\tanh A_{s}, \label{A3} \\
-(K-p_{s}) \v_{s} &=& p_{s}^{2}-4\eta_{s}^{2}-\delta\pm \nonumber \\ 
\,\ &\,& \frac{a \pi}{2 \eta_{s}} 
A_{s}\, 
\mathrm{sech} A_{s} \tanh A_{s}, \label{A4}
\end{eqnarray}
where $A_{s}=\pi (K-p_{s})/(4 \eta_{s})$. 
 For either sign, the system (\ref{A3})-(\ref{A4}) has a single root 
$\v_{s}=2 p_{s}$, $p_{s}=K$ and 
$\eta_{s}=\frac{1}{2} \sqrt{K^{2}-\delta}$. Therefore, there are two fixed points 
on the real axis, located at $\mathrm{Re} \Psi = \pm \sqrt{K^{2}-\delta}$. 

For the set of parameters of Fig. \ref{fig2} there are a stable and 
an unstable  
fixed point close to $+1$ and $-1$, respectively. The stability of the 
fixed points is 
determined by solving the collective coordinates equations numerically  
with the initial conditions very close to the above values, e.g., 
$\eta_{0}=\eta_{s}+10^{-8}$, $q_{0}=0$, $p_{0}=p_{s}$, 
$\Phi_{0}=\Phi_{s}^{\pm}=\pm \pi/2$. 
In the unstable case ($\Phi_{0}=-\pi/2$), 
 the numerical solution exhibits oscillations of the amplitude and  
 phase, whereas the velocity of the soliton remains constant. 
This solution is 
represented by the separatrix in Fig.\ \ref{fig2}. 
A trivial stable fixed point is located at the origin; its stability is 
established by numerical solutions of the collective coordinates equations with 
$\eta_{s}$ close to zero.

\end{document}